# Study of stability of nuclei, flow and multifragmentation in heavy-ion collisions.

Supriya Goyal*
*Department of Physics, Panjab University, Chandigarh-160014, INDIA*
*\* email: supriyagoyal.pu@gmail.com*

## Introduction and Methodology

The aim of the present thesis is to theoretically study the multifragmentation as well as its related phenomena like nuclear flow, for hot and dense nuclear matter formed in the symmetric as well as asymmetric heavy-ion collisions at intermediate energies. The Quantum Molecular Dynamics (QMD) [1] model is used for to generate the phase-space of nucleons. The phase-space is then, analyzed using different clusterization algorithms. The QMD model is based on a molecular dynamic picture where nucleons interact via two and three-body interactions. The QMD approach treats the nucleons as Gaussian wave packets. The nucleons propagate according to the classical equations of motion:

$$\frac{d\mathbf{r}_i}{dt} = \frac{dH}{d\mathbf{p}_i} \text{ and } \frac{d\mathbf{p}_i}{dt} = -\frac{dH}{d\mathbf{r}_i}, \qquad (1)$$

where H stands for the Hamiltonian which is given by

$$H = \sum_i \frac{\mathbf{p}_i^2}{2m_i} + V^{tot}. \qquad (2)$$

Our total interaction potential $V^{tot}$ reads as

$$V^{tot} = V^{Loc} + V^{Yuk} + V^{Coul} + V^{MDI}, \qquad (3)$$

where $V^{Loc}$, $V^{Yuk}$, $V^{Coul}$, and $V^{MDI}$ are, respectively, the local (two and three-body) Skyrme, Yukawa, Coulomb and momentum dependent potentials.

## Results and discussion

As the first part of the thesis, we propose an extension of the spatial clusterization algorithm i.e. minimum spanning tree (MST) algorithm by incorporating microscopic binding energy calculated from the Bethe Weizsäcker Mass (BWM) formula [2]. The extension is termed as MSTB(2.1). It is found that present implementation of MST method is promising as it avoids creation and detection of spurious fragments and, in many cases, yields results close to SACA(2.1) method and to experimental data of symmetric reactions [3]. We further extend the study on the mass asymmetric systems [4]. The asymmetry of a reaction is defined by the parameter called asymmetry parameter ($\eta$) and is given by:

$$\eta = \left|\frac{A_T - A_P}{A_T + A_P}\right|, \qquad (6)$$

where $A_T$ and $A_P$ are the masses of target and projectile, respectively. The $\eta = 0$ corresponds to the symmetric reactions and nonzero values of $\eta$ defines different asymmetries of a reaction. We analyze the role of mass asymmetry of the reaction on the multifragmentation by varying the mass asymmetry from 0 to 0.74. A significant role of the mass asymmetry on the dynamics of the reaction is seen using MSTB(2.1) and SACA(2.1) clusterization algorithms. The multiplicities of various fragments are found to decrease with increase in the mass asymmetry of the reactions. A comparison with experimental data is also done.

A detailed study of the energy of vanishing flow with reference to mass asymmetry is also presented [5]. Almost independent of the system mass as well as impact parameter, a uniform effect of the mass asymmetry is seen on the energy of vanishing flow (EVF). We find that for large asymmetries ($\eta$=0.7), the effect of asymmetry is 15% with MDI and in the absence it is nearly 40%. This also explained the deviation in the individual EVF from the mean values as reported in earlier studies. We also investigate the role of colliding geometry in the disappearance of flow as well as its mass dependence throughout the mass range 40-240 for mass-asymmetric reactions with η = 0.1-0.7. We find that the effect of mass asymmetry is more dominant at peripheral collisions than at central ones. Very interestingly, we find that the percentage change in EVF with colliding geometry remains uniform throughout the mass asymmetry range for every fixed system mass. At a constant energy of 200 MeV/nucleon, we also find that the geometry of vanishing flow (GVF) is quite sensitive to the mass asymmetry. For each η, the mass dependence of GVF follows a power law behaviour and dependence on the system mass increases with increase in η.

The second last problem is devoted to a systematic study of the formation of fragments with different mass ranges in collisions of Au+Au at incident energies between 20-1000 MeV/nucleon and at impact parameter between 0 and $b_{max}$ [6]. The aim of present study is to understand the complex dependence of fragment production on incident energy and impact parameter. For better understanding, the results with different clusterization algorithms are also shown. Our results clearly indicate that (i) a rise and fall in the fragment production for central and semi-central collisions with an increase in the incident energy. The peak in the multiplicity occurs at higher incident energies for light mass fragments compared to heavy mass fragments. This needs experimental verification. (ii) the IMF production as a function of impact parameter exhibits well established behaviour i.e. the multiplicity is maximum for central collisions and it decreases with the increase in impact parameter at low incident energies, whereas it has a clear rise and fall at higher incident energy. Quite similar behaviour is seen with all clusterization algorithms (iii) no signal of liquid-gas phase transition is seen as no unique dependence of τ on impact parameter is seen. Moreover, the incident energy at which τ has minima is also not same at all impact parameters and clusterization algorithms. (iv) no rise and fall in the fragments with A = 2 and LCP's (2≤A≤4) with change in impact parameter at all incident energies. While the fragments with mass A = 4, 5≤A≤9, 5≤A≤20, IMF's, and HMF's show a rise and fall for E > (400,400,250), (250,250,250), (250,250,150), (150,150,150), and (100,100,100) MeV/nucleon using MST,MSTB(2.1),SACA(2.1), respectively. The energy of onset of multifragmentation and vaporization varies with the clusterization algorithms and is found to be 60 MeV/nucleon and 150 MeV/nucleon with SACA(2.1). In other words, the QMD simulations of $^{197}Au+^{197}Au$ predict different behaviour for different mass ranges than for IMF's with a change in incident energy, impact parameter, and clusterization algorithms.

The role of different widths of Gaussian wave packets (i.e., L = 1.08 $fm^2$ and L = 2.16 $fm^2$) on the stability of the ground state of nuclei through out the periodic table for different equations of state within QMD model [7]. We find that the broader Gaussian wave packet yields stable nuclei which emit very few spurious nucleons. This observation is more valid for heavy mass nuclei where proper ground state properties such as binding energy, density distribution, and root-mean-square radii can be obtained.